%% file: main.tex
\documentclass[%
 reprint, % for twocolumn
groupedaddress,
 preprintnumbers,
 showkeys,
 amsmath,amssymb,
 aps,
]{revtex4-2}
\usepackage{url}
\usepackage[inline]{enumitem}
\usepackage{wrapfig}
\usepackage{graphicx}
\usepackage{float}
\usepackage[titles,subfigure]{tocloft}
\floatstyle{boxed}
\usepackage[margin=1.0in]{geometry}
\usepackage{setspace}
\usepackage{fancybox}
\usepackage{booktabs}
\usepackage{hyperref}
\usepackage{caption}
\usepackage{subcaption}
\usepackage{mathrsfs}
\usepackage{tablefootnote}
\usepackage{xcolor} %Needed for the individual highlighting below
\usepackage{soul}
\usepackage[ruled,vlined]{algorithm2e}
\graphicspath{Figures}
\usepackage{scalerel}
\usepackage{tikz}
\usetikzlibrary{svg.path}

\definecolor{orcidlogocol}{HTML}{A6CE39}
\tikzset{
  orcidlogo/.pic={
    \fill[orcidlogocol] svg{M256,128c0,70.7-57.3,128-128,128C57.3,256,0,198.7,0,128C0,57.3,57.3,0,128,0C198.7,0,256,57.3,256,128z};
    \fill[white] svg{M86.3,186.2H70.9V79.1h15.4v48.4V186.2z}
                 svg{M108.9,79.1h41.6c39.6,0,57,28.3,57,53.6c0,27.5-21.5,53.6-56.8,53.6h-41.8V79.1z M124.3,172.4h24.5c34.9,0,42.9-26.5,42.9-39.7c0-21.5-13.7-39.7-43.7-39.7h-23.7V172.4z}
                 svg{M88.7,56.8c0,5.5-4.5,10.1-10.1,10.1c-5.6,0-10.1-4.6-10.1-10.1c0-5.6,4.5-10.1,10.1-10.1C84.2,46.7,88.7,51.3,88.7,56.8z};
  }
}
\newcommand\orcidicon[1]{\href{https://orcid.org/#1}{\mbox{\scalerel*{
\begin{tikzpicture}[yscale=-1,transform shape]
\pic{orcidlogo};
\end{tikzpicture}
}{0}}}}
\newcommand{\PUNT}[1]{} 

\input{custom_macros}
\begin{document}

\pagestyle{empty}
\pagestyle{plain}
\pagenumbering{arabic}
\setcounter{page}{1}

%\begin{frontmatter}

\title{Uncertainty aware anomaly detection to predict errant beam pulses in the SNS accelerator }

\author{Willem Blokland\,\orcidicon{0000-0002-2385-6894}}
\author{Pradeep Ramuhalli\,\orcidicon{0000-0001-6372-1743}}
\author{Charles Peters}
\author{Yigit Yucesan\,\orcidicon{0000-0002-2627-0821}}
\author{Alexander Zhukov\,\orcidicon{0000-0002-1363-9343}}
%\author{Yigit}
\affiliation{Oak Ridge National Laboratory, Oak Ridge, TN 37830, USA }

\author{Malachi~Schram\,\orcidicon{0000-0002-3475-2871}}
\author{Kishansingh Rajput\,\orcidicon{0000-0002-4430-9937}}
\author{Torri Jeske\,\orcidicon{0000-0002-5191-0501}}
\affiliation{Thomas Jefferson National Accelerator Facility, Newport News, VA 23606, USA}

% Force line breaks with \\
\date{\today}% It is always \today, today,

\begin{abstract}
High-power particle accelerators are complex machines with thousands of pieces of equipment that are frequently running at the cutting edge of technology. In order to improve the day-to-day operations and maximize the delivery of the science, new analytical techniques are being explored for anomaly detection, classification, and  prognostications. As such, we describe the application of an uncertainty aware Machine Learning method, the Siamese neural network model, to predict upcoming errant beam pulses using the data from a single monitoring device. By predicting the upcoming failure, we can stop the accelerator before damage occurs. We describe the accelerator operation, related Machine Learning research, the prediction performance required to abort beam while maintaining operations, the monitoring device and its data, and the Siamese method and its results. These results show that the researched method can be applied to improve accelerator operations. 
\end{abstract}
\keywords{accelerator, beam current, anomaly prediction, machine learning, Siamese neural network}
\maketitle

%{\let\thefootnote\relax\footnote
%\end{frontmatter}
\input{Sections/Intro.tex}

\input{Sections/Methods.tex}
\input{Sections/Results.tex}

\input{Sections/Discussion.tex}

\bibliographystyle{unsrt}
\bibliography{References/PR.bib,References/MS.bib}
%\bibliography{References/MS.bib}

\end{document}

%% file: custom_macros.tex
\newif\iffinal

\definecolor{limeyellow}{RGB}{228,255,100}
\usepackage[colorinlistoftodos,prependcaption,textsize=footnotesize,color=limeyellow,linecolor=magenta]{todonotes}

\iffinal
    
    \newcommand{\ar}[1]{}
    \newcommand{\fyl}[1]{}

\else
    \newcommand{\ar}[1]{\todo[inline,linecolor=blue,backgroundcolor=blue!25,bordercolor=blue,textcolor=black]{#1}}

    \newcommand{\fyl}[1]{\ar{[FYL] #1}}

\fi

%% file: Sections/Intro.tex
\section{Introduction}

The Spallation Neutron Source (SNS) facility is the world’s highest power proton accelerator, delivering 1.4 MW of a 1 GeV pulsed beam at 60 Hz. The beam is accelerated in the linear accelerator which has both a warm, normal conducting and a cold, superconducting section. The accelerated beam is injected into the accumulator to form a very short but intense pulse of intensities up to $1.4 \times 10^{14}$ protons per pulse that is sent to a stainless steel vessel filled with liquid mercury \cite{2014Henderson-SNSSystemDesign}. The impact of the protons spalls the mercury atoms and neutrons are released. These neutrons are then guided to experimental beam lines where the material research takes place. 

Achieving high availability is extremely difficult in high-power proton beam accelerators. These accelerators use thousands of subsystems, with many running on the cutting edge of technology. Errant beam pulses can cause damage to the accelerator and negatively impact the research program. To minimize down times, accelerator operations include preemptively replacing equipment, careful scheduling of maintenance periods, utilizing diagnostic instruments to equipment, and detailed tracking of downtime statistics and patterning. These measures have limitations as failures still happen unexpectedly. Adding more diagnostics instruments could help but is be very expensive. 

There is therefore a need for methods that can utilize existing diagnostic data to identify the onset of errant beam pulses. Based on years of analysis of beam trip data, errant beam pulses are caused by equipment failure and as nearly all equipment involved in the acceleration process must have an effect on the beam, we assume that conditions leading to errant beam pulses can be identified by monitoring signals from beam measurements.

This paper describes the results of research being conducted to exploit the extraordinary advantages of machine learning (ML) and the vast amounts of accelerator data in a neutron production facility to improve accelerator availability. The focus of this research is on using ML to predict beam loss due to the failure of various accelerator equipment, using data from a single existing beam monitoring device. If successful, we can avoid the cost of having to upgrade or install additional monitoring devices to predict upcoming failures.

\section{Previous Work}

%~\cite{PhysRevAccelBeams.23.114601-Tennant}...References and overview (see PR.bib)....  

To date, ML has been used in a limited capacity in the accelerator and target community and has primarily focused on improvements in beam tuning and beam quality \cite{2019-Leeman-MLStabilizationSynchrotron,2018-Emma-MLPhaseSpacePrediction}. 

The focus of recent studies are on improving accelerator operations by detecting deviations from normal conditions. To this end, several approaches have been considered. Relevant literature focuses on building models of the beam behavior in order to identify scenarios that correspond to beam or equipment errors. 
Fol et al \cite{2018-Fol-MLBeamDiag} provide an overview of several potential applications of ML for accelerators, including optimization and prediction for tuning accelerator operations, lattice imperfection corrections, and anomaly detection. In addition to this high level overview, two specific applications of ML at the Large Hadron Collider (LHC) are discussed: optics correction (predicting control knob setting to cancel quadrupole field errors) and anomaly detection (detecting faulty BPM)
 \cite{2019-Fol-ML4BPMFaults}. In both cases, the proposed ML produce reasonable results, though the results described seem to show room for improvement in terms of True Positive (TP) and False Positive (FP) rates. 
Furthermore, Fol et al \cite{PhysRevAccelBeams.23.102805} utilize Isolation Forest (IF) method to identify faulty BPM data that cannot be detected with singular value decomposition (SVD) method. Results of the study illustrate that $~90\%$ of unidentified bad BPM data can be isolated when SVD and IF methods are applied together. Emma et al \cite{2018-Emma-MLPhaseSpacePrediction} describe an MLP-based prediction of the longitudinal phase space of particle accelerators based on diagnostic measurements. Results from simulation and data from the LCLS indicate good prediction accuracy (mean prediction accuracy ranging from $~$0.6 to $~$0.85 depending on the data set used for the training).  Similarly, \cite{2018-Valentino-MLBeamLossLHC} discuss a gradient boosting classifier for identifying beam loss plane contributions to the measured Beam Loss Monitor (BLM) data at the LHC. Other studies \cite{PhysRevAccelBeams.23.114601-Tennant} discuss the application of ML for Superconducting Radio Frequency (SRF) cavity fault classification. In this study, Radio-Frequency(RF) waveforms from fault conditions are used to assess the potential of different ML algorithms for the classification task. Reported results indicate the ability to accurately (85\% accuracy) identify the cavity and cavity fault mode.  Wielgosz and Skocze{\'{n}} \cite{Wielgosz2019} propose using a Long Short-Term Memory (LSTM) cell to model superconducting magnets in the LHC in order to imitate operational time series data. After ensuring the ability of the model to represent the system accurately, the authors test the anomaly detection capabilities of their approach. Results indicate that the anomalous behavior can be captured through deviation of data from model predictions. Sanchez-Gonzalez et al \cite{SanchezGonzalez2017} undertake prediction of shot-to-shot X-ray properties in an X-ray Free-Electron Laser (XFEL) using standard machine learning tools such as artificial neural networks and support vector regression. Presented performance metrics indeed prove that high agreement (as high as 97\% on specific variables) between predictions and measured data can be achieved using basic ML models on high-repetition rate data, which allows more accurate and preemptive diagnostics for particle accelerators.

While such anomaly detection techniques are promising, there are few studies using data from a limited number of beam monitors to identify conditions leading to beam loss. The goal in this paper is to predict the onset of conditions leading to errant beam pulses at least one pulse in advance from beam measurements. Similar questions are studied in \cite{2020-Rescic-AccelFailure} and \cite{2021-Li-AccelTimeSeries}. Rescic et al \cite{2020-Rescic-AccelFailure} demonstrate that a Random Forest (RF) is capable of identifying a beam loss event at SNS one pulse in advance using data from one beam current monitor. In \cite{2021-Li-AccelTimeSeries}, the authors examine the potential to predict interlocks (reflecting beam shutoff) using multiple measurements along the accelerator. A recurrence-plots based convolutional neural network (RPCNN) is used to convert the measurements into recurrence plots that are then classified using a convolutional neural network (CNN). Comparisons with a RF approach indicate that the performance of the RF and RPCNN are comparable, though the RPCNN is more successful at identifying anomalies that build up over time. The RPCNN can identify the onset of faults earlier than the RF, although the approach requires data from multiple measurements and may take additional computational time to generate the recurrence plots. It is worth noting that both solutions are challenged when faced with beam loss events that are dissimilar to those in the training data set.

Here, we propose to use a Siamese Neural Network model to provide a similarity ranking between a normal and an anomalous pulse in order to predict an upcoming errant beam pulse. We utilize a Differential Current Monitor (DCM) that uses data from two beam current sensors located upstream and downstream of the source. The sensors used in this study are the same as those in \cite{2020-Rescic-AccelFailure}. Our approach specifically addresses the use of a limited amount of monitoring devices to identify beam loss events, avoids extensive computational loads and the use of data from multiple measurements, can adjust the threshold to change the FP and TP rates, ca determine is the trained model is outdated, and is not affected by anomalies the model is not trained on.

\section{Accelerator performance}
To evaluate whether our ML method can be applied in practice, we must use a figure of merit for accelerator performance expressed in ML terms.  One of the key metrics to track the performance of the SNS facility is the beam availability.  The beam availability is defined as the number of beam hours delivered versus the number of beam hours scheduled. Ideally, the beam availability should be $100 \%$ but in practice the maximum achievable yearly beam availability for the facility is currently $95 \%$. The reduction in beam availability is tracked closely by recording beam trip causes, frequencies, and durations.  To prevent a negative impact on the scientific research, any method that preemptively and wrongly aborts the accelerator beam should not noticeably increase the current levels of beam trip frequency or duration. While many machine learning applications often allow single digit percentages of FP rates, in the SNS case, a 5\% FP level would reduce our integrated power on target by 5\%, which would adversely affect the material science research program. We will define an acceptable FP rate by analyzing the number of beam trips and trip durations. 
\subsection{Classes of errant beam}
In the context of this paper, there are two categories of errant beam pulses, those where the beam pulse is aborted and thus shortened but does not cause significant beam loss in the Superconducting Cavity Linac (SCL), we label those 1100 events, and those that do cause significant beam loss in the SCL, labeled as 1111 events. The first two digits reflect beam truncation while the last two digits reflect beam loss. It is currently difficult to find a direct correlation between non beam loss errant beam pulses and SCL degradation, but there are clear correlations between beam loss and SCL degradation. 

\begin{figure}[ht]
\centering
\includegraphics[width=0.475\textwidth]{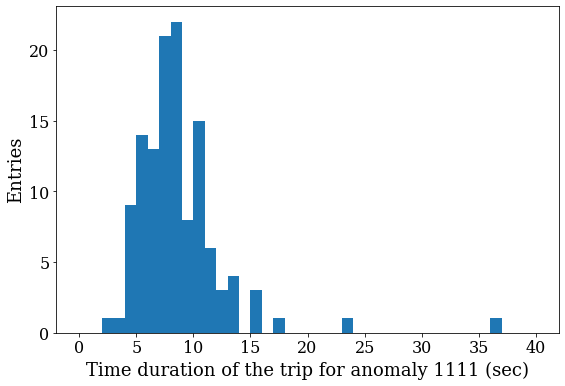}
\caption{Histogram of the duration between the errant beam pulse and the next available beam pulse for 1111 events (beam truncation with SCL beam loss).}
\label{fig:Hist1111}
\end{figure}

When the errant beam pulse with beam loss is aborted, the accelerator does not automatically restart and an operator has to manually intervene to restart the accelerator. This setup allows the operator to evaluate whether the beam can be turned back on without causing additional beam loss in the accelerator. For the analyzed data-set for the month of March 2021, the total beam production time was $\approx$26.4 days with average daily trip frequency of SCL beam loss trips of $\approx$5 per day with an average beam off time of $\approx$9 seconds, see Figure \ref{fig:Hist1111}. Not included in the recovery time is a required 30 second beam power ramp to full beam power which then indicates an end to the downtime. Including the 30 second recovery there were then 309,000 total pulses lost per day or 0.22\%. 
Though for this particular month the number of pulses missed from beam loss events was low previous experience has identified that the beam loss can trigger SRF cavity degradation. For a recent beam loss event in July 2021 an SRF issue was triggered by beam loss from an upstream RF cavity fault, and the beam recovery for that particular event took $\approx$1 hour. Not only did a long down-time result, but SRF cavity gradients were reduced to maintain high reliability. Such long down-times are more troublesome to experimenters as they can lose a significant amount of their allotted beam-time while the lower cavity gradients might require a long maintenance period to treat the cavities and recover their performance. Avoiding pulses beam loss can thus not just avoid the less than one minute down-times but also the rarer, longer down-times.

\begin{figure}[ht]
\centering
\includegraphics[width=0.475\textwidth]{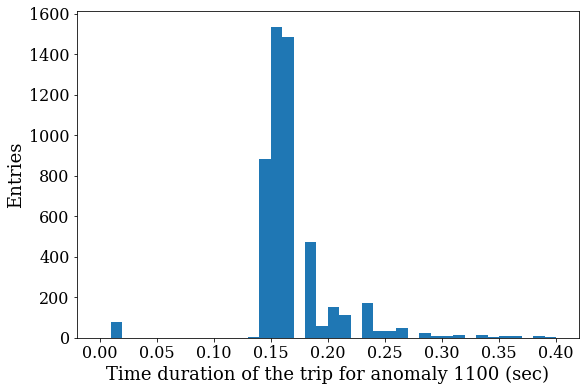}
\caption{Histogram of the duration between the errant beam pulse and the next available beam pulse for 1100 events (beam truncation without SCL beam loss).}
\label{fig:Hist1100}
\end{figure}

The 1100 events can auto restart in as little as 4 pulses. Figure \ref{fig:Hist1100} shows the histogram of 1100 events for the month of March in 2021 that are shorter than .4 seconds to illustrate the auto restart times. Interestingly, this figure does show a longer duration between the errant beam and the first pulse after restart, 8 pulses (pulses are 16.6 ms apart). This is due to an inadvertent abort system setting for this period but this can be shortened to 4 pulses. Many of the shorter 1100 events are due to dropouts of the Radio Frequency Quadrupole (RFQ) cavity, about 5500 over the data-set. This equates to $\approx$200 per day, or 1,600 pulses per day or 0.03\% of beam. Note that these, when spread out, are not noticeable in the beam power as these are typically spread out over minutes and within the noise of the beam power measurements and thus have not been part of the accelerator metrics. While most of the truncated beam pulses have no further effects, it is possible that problems occurs in a cavity that was loaded for a longer beam pulse. The fields will temporarily increase beyond the intended maximum strength exponentially increase field emitted electrons which could cause cavity problems and longer down-times. We have not analyzed the cause of the longer down-times of the 1100 events, the overall downtime associated is 30,000 seconds over the month or 1.3\%. Not all of this can be prevented by aborting beam before the errant beam pulse. For example, 2.5 hours of this down-time was caused by the Personnel Protect System which is unrelated to the acceleration process. 

\subsection{Acceptable lost pulses}
When focusing on evaluating the benefit of the ML implementation in reducing the frequency of beam loss events in the SCL, we deal with an order of 0.22\% of beam lost. A TP rate of 0.5 would gain over 0.1\% of beam. The ML implementation should predict this type of event and inform the Machine Protection System (MPS) to not send a beam pulse until the fault condition has cleared. This clearing of the fault could be very quick, assuming a glitch in the equipment, and we can reissue beam after the minimal 4 pulses hold-off. This type of fault would then be an auto restart fault instead of the longer manual reset. If 
we maintain our erroneously aborted beam pulses lower than 0.1\% we would not even have additionally lost additional pulses and protect our accelerator from damage due to beam loss.
While the truncated beam pulse due to the RFQ drop-outs might be predicted, the cost of avoiding is the same as the actual abort. Still, the ML implementation could potentially prevent the longer down-times accounting for about 1\% of beam lost. 

We will not know the actual benefit of the ML implementation until it is activated, but given the current tolerated down-times, we will aim for a FP rate resulting in loss of around 0.2\% of beam with a TP rate for 1111 events of 50\% while acknowledging that slightly larger or smaller FP rates are a trade-off between losing beam pulses but minimizing damage to the accelerator and interruptions to the research program. Given the 4 pulses downtime per abort, we need a FP rate of 0.05\%. This would put our recall, $TP/(TP+FN)$, at 0.5 and our precision, $TP/(TP+FP)$, at 0.999.

\section{Overview of Differential Current Monitor}

\begin{figure}[h]
\centering
\includegraphics[width=0.475\textwidth]{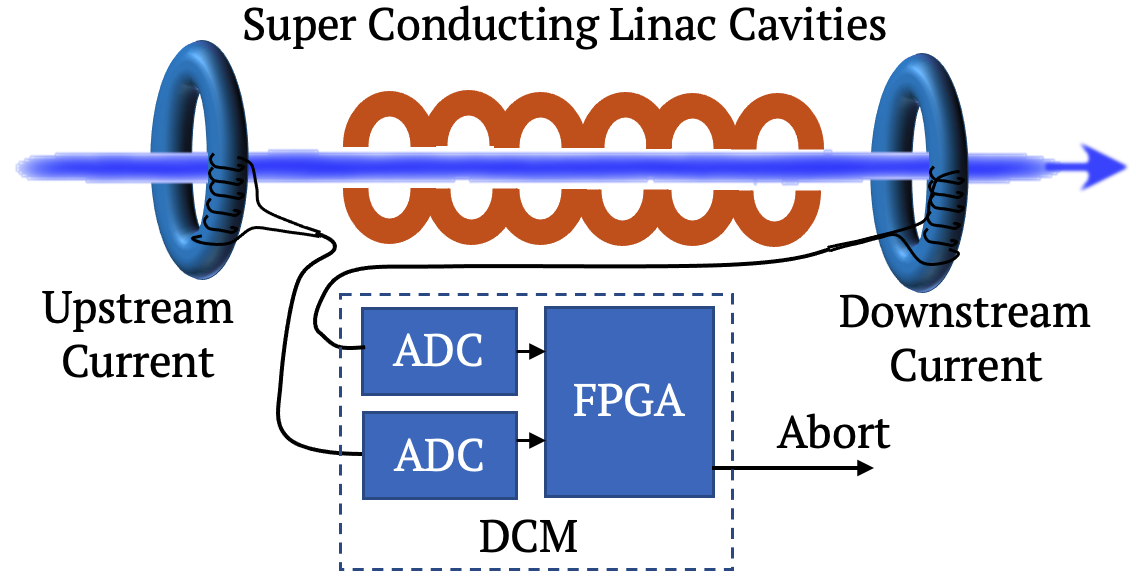}
\caption{Setup of the Differential Current Monitor showing one mini-pulses as it passes through the SCL.}
\label{fig:sns-dcm}
\end{figure}

The beam monitoring device providing the data is the Differential Current Monitor (DCM) \cite{dcm-1.1, dcm-2.2}. This device monitors the beam current upstream and downstream of the super conducting section of the accelerator. It aborts beam when it detects a difference between the upstream and downstream beam current, the 1111 event. The DCM can also abort when there is a significant pulse-to-pulse difference, the 1100 event. This form of errant beam must also be aborted, as the cavities' controller assumes the next pulse is also of the shorter length and will not output enough power for the next pulse, resulting in beam losses. The DCM can abort for this scenario and this will tell the cavity controller to not learn on the aborted pulse and keep the settings for the normal pulse.  The DCM setup is shown in Figure \ref{fig:sns-dcm}. The two beam current signals are fed into the Field Programmable Gate Array (FPGA) for comparison and generation of the abort signal when the difference exceeds a threshold. The difference signal is fed into two integrating sliding windows, each having its own window length and threshold. The shorter window is setup to detect sudden large losses and the longer window to detect gradual losses.

\begin{figure}[ht]
\centering
\includegraphics[width=0.44\textwidth]{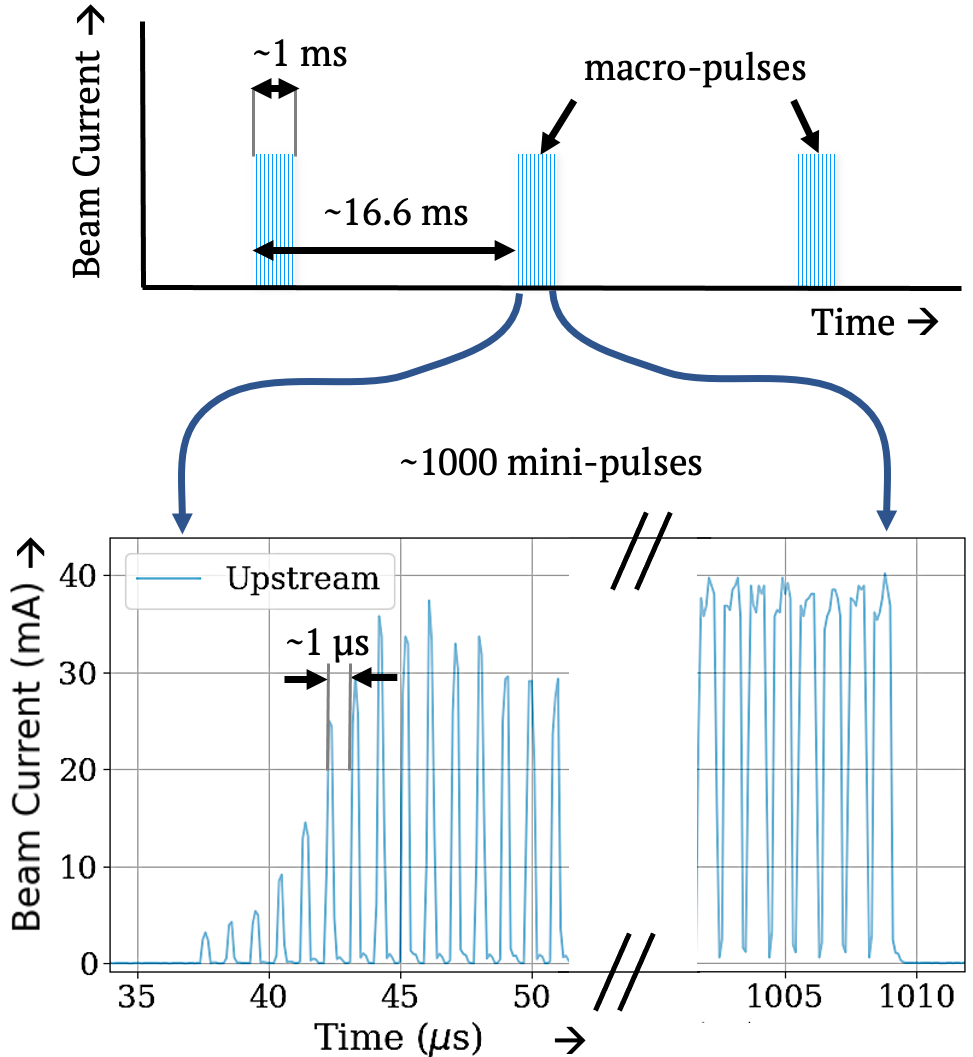}
\caption{Pulse pattern and beam current trace.}
\label{fig:sns-waveforms4}
\end{figure}

The pattern of the beam current is shown in Figure \ref{fig:sns-waveforms4}. 
The upper plot shows what is referred to as a series of macro-pulses, a 1 ms long pulse repeated at 60 Hz. This macro-pulse consists of approximately 1000 mini-pulses. Each mini-pulse is \~650 ns and is followed by a gap of 350 ns. Later in the accelerator, these mini-pulses are accumulated in the ring and stacked on top of each other to create a short, 650 ns high intensity pulse that is directed to the target to produce the neutrons. The bottom plot shows an actual measured beam current signal with the initial ramp-up in intensity in the beginning of the macro-pulse as well as the different widths of the mini-pulses during the macro-pulse. This setup is typical for production style beam.

While other devices such as BLMs can also abort beam, the DCM is quicker, up to three times, to abort the beam by virtue of its fast processing in its FPGA and its special, dedicated connection to the abort system. Another unique feature of the DCM, one that makes it possible to use it for ML, is that it not only archives the beam current traces during the errant beam pulse but also the beam current traces before the errant beam pulse, and a normal beam trace on a regular schedule. The traces can now be used for (semi)-supervised learning with the before trace of an errant beam pulse event as the anomalous class and the before of the normal trace as the normal class.

%% file: Sections/Methods.tex
\section{Methods}\label{sec:Methods}

\subsection{Siamese Model}\label{sec:SiameseModel}
\begin{figure}[ht]
    \centering
    \includegraphics[width=0.45\textwidth]{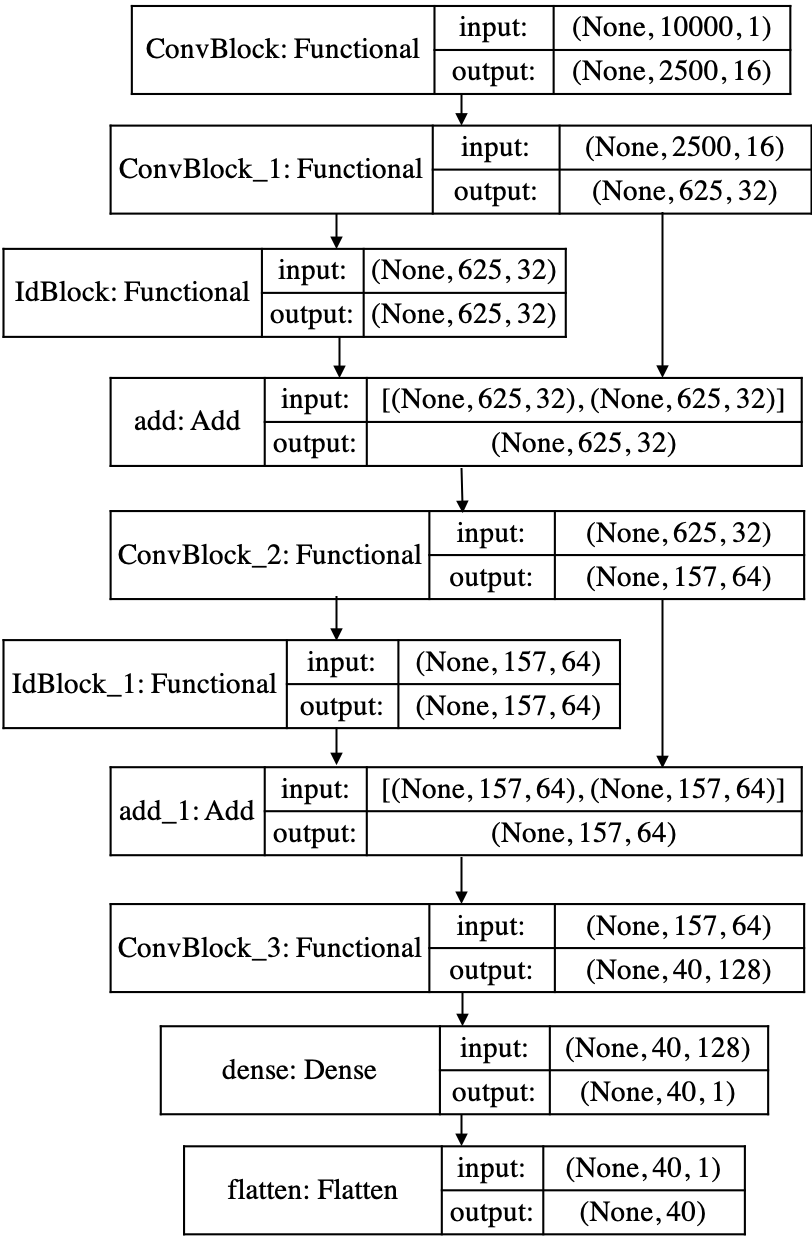}
    \caption{ResNet architecture used in our Siamese model}
    \label{fig:ResNetPlot}
\end{figure}

For this study, we explore the use of Siamese neural network models~\cite{Koch2015SiameseNN} to provide a natural similarity ranking between two inputs.  
Our aim is to use the Siamese model to learn the similarities between normal and anomalous pulses, as measured by the SNS DCM sensors.
Developing a machine learning model based on a similarity score provides robustness against previously unseen anomalies that could be introduced to the system.
Additionally, the similarity score can be used to re-evaluate the applicability of the current model by comparing samples used for training with the current normal pulse. 
A trending change in the similarity score would indicate the need to re-train the current Siamese model.\\ \\  
A Siamese model consists of twin networks that accept unique inputs.
The twin networks are used to shrink the large raw data input to a reduced representation that captures the salient features.
The reduced representations of each input are then compared using a modified contrastive loss function~\cite{1640964}:
\begin{equation}
L(y,\hat{y}) =    \alpha \times (1-y)*\hat{y}^2 + y * (\max(\beta-\hat{y},0))^2
\label{eq:contrastive_loss}
\end{equation}
The contrastive loss function is composed of two terms used to decrease the output of like pairs and increase the output of unlike pairs. 
Here $y$ is the truth value, $\hat{y}$ is the predicted value, $\alpha$ is tuning parameter use to emphasize the similar pulses that was set to 2 for this study, and $\beta$ is a second tuning parameter, set to 1, used to emphasize dissimilar pulse. 
%however, there are a several other loss functions have been proposed for Siamese networks~\cite{}
We used a ResNets~\cite{He2016DeepRL} model for the twin network. 
% \KISHAN{Provide summary of the explicit implementation for our paper\\}
ResNets consist of several stacked residual units, which can be thought of as a collection of convolutional layers coupled with a ‘shortcut’ that improves the propagation of the signal in a neural network. 
This ‘shortcut’ allows for the construction of much deeper networks, since keeping a ‘clean’ information path in the network facilitates optimization. 
The architecture of the ResNet model is shown in Figure \ref{fig:ResNetPlot}.

\begin{figure}[ht]
    \centering
    \includegraphics[width=0.475\textwidth]{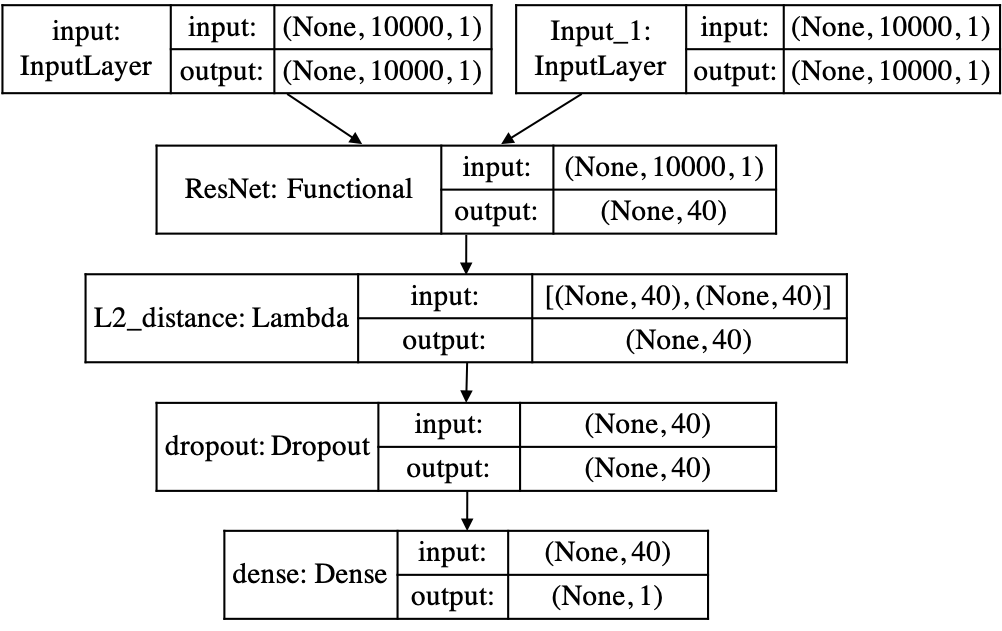}
    \caption{Siamese model architecture}
    \label{fig:SiamesePlot}
\end{figure}

\begin{figure}[ht]
    \centering
    \includegraphics[width=0.475\textwidth]{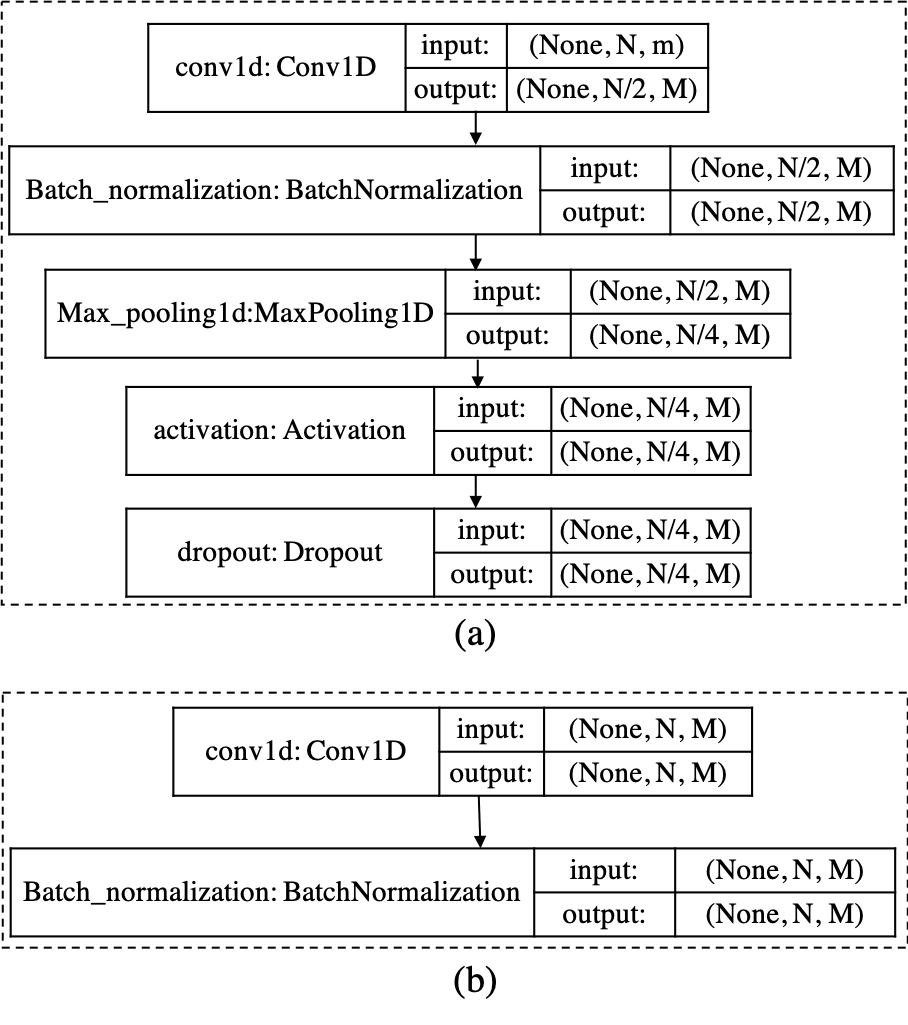}
    \caption{Architecture of (a) ConvBlocks, and (b) IdBlocks used in ResNet}
    \label{fig:conv_block-id_block}
\end{figure}

The model was developed using \textsc{Keras}~\cite{keras} and Tensorflow back-end~\cite{tensorflow2015-whitepaper}.
We used the Adam optimizer~\cite{adam} and a cost function as defined in Equation~\ref{eq:contrastive_loss}. 
For our study, we used the similarity metric as defined in Equation~\ref{eq:l2_metric}:
\begin{equation}
L^{2} = \left| \sum_{i=0}^{N} (x_{1,i}^{2} - x_{2,i}^{2}) \right| 
%
%L^{2} = \left| \sum_{i=0}^{N} (x_{1}^{i})^{2} - (x_{2}^{i})^{2} \right|
%
%L^{2} = \left| \sum_{i=0}^{N} (x_{1}(i)^{2} - x_{2}(i)^{2}) \right|
\label{eq:l2_metric}
\end{equation}
Here $x_{1}$ and $x_{2}$ are the latent vector outputs from the ResNet model for input pulse 1 and 2, and $i$ is the element wise index. 
%\begin{ruledtabular}
%\begin{tabular}{r|c|c|c|c}
%Layer & Layer Type & Filters & Strides & Activation \\
%\hline
%1 & Conv1D & 16 & 2 & $\tanh$  \\
%1 & Conv1D & 32 & 2 & $\tanh$ \\
%1 & Conv1D & 64 & 2 & $\tanh$  \\
%1 & Conv1D & 128 & 2 &$\tanh$ \\
%1 & Dense & 1 & $\tanh$  \\
%\end{tabular}
%\end{ruledtabular}
%\end{table}
\subsection{Uncertainty Aware Siamese Model }

Providing methods to reliably quantify the predictive uncertainty for our models is critical for real-world applications.
This is acutely visible when the input samples are dissimilar to the training sample.
The use of distance-awareness is particularly important because deterministic models are only trained on a data set $\mathscr{D}=\{y_{i},\mathbf{x}_{i}\}^{N}_{i=1}$ where $\mathscr{D}$ is a subset of the input space, $\mathscr{H}_{IDD} \subset \mathscr{H}$. 
Consequently, the model only learns the in-domain distributions, $p^{*}(y|\mathbf{x},\mathbf{x} \in H_{IDD})$ from the data set $\mathscr{D}$ and there is a possibility of an orthogonal data set, $p^{*}(y|\mathbf{x},\mathbf{x}\notin \mathscr{H}_{IDD})$ existing in the same input space. 
This orthogonal distribution $p^{*}(y|\mathbf{x},\mathbf{x}\notin \mathscr{H}_{IDD})$ can be very different from the training distribution, the predictions on this orthogonal set can lead to bad results.
Gaussian Processes are able to provide predictions with uncertainties, however, this techniques does not scale for large data samples and high dimensional problems.  
% For this study, we used the method proposed in paper~\cite{} the to provide an approximation of the 
For this study, we extend the Siamese model described in Section~\ref{sec:SiameseModel} by replacing the output layer with a Gaussian Process (GP) as described in~\cite{liu2020simple}. 
A classic deep learning model maps the input space to a hidden representation space and it's output layer maps the hidden representation $h(x)$ to the label space $y$. 
By wrapping a GP layer around the output layer, we make it distance aware such that it outputs an uncertainty score representing distance between the hidden space of the test data to that of the input space (distribution that the model is trained on) $|h(x) - h(x')|$.

%% file: Sections/Results.tex
\section{Results}
% The machine learning implementation was done using tensorflow~\cite{tensorflow2015-whitepaper}
% The input pulses was converted to a one dimensional convolution to enable the use of the ResNET architecture.
% The v8 10k samples 
For this study, we used the Receiver Operating Characteristic (ROC) curve to quantify the performance of our models.
The ROC curve indicates the relationship between true positives (TP) and false positives (FP).
In our case, a true positive is defined as correctly identifying an anomaly and a false positive is defined as incorrectly identifying a normal pulse as anomalous.
We trained two Siamese models with identical architectures/configuration with the exception of the output layer as explained in Section~\ref{sec:Methods}.
%For the deterministic model, we use a dense layer and fir the distance aware model we wrapped the dense layer with a Gaussian processes layer.
%(deterministic model) and another with Gaussian processes. 
% We set $\alpha$ term for the contrastive loss function to 2 to assist the model in learning the similarity between normal traces and to lower the amount of false positives. 
\begin{figure}[b]
\centering
\includegraphics[width=0.475\textwidth]{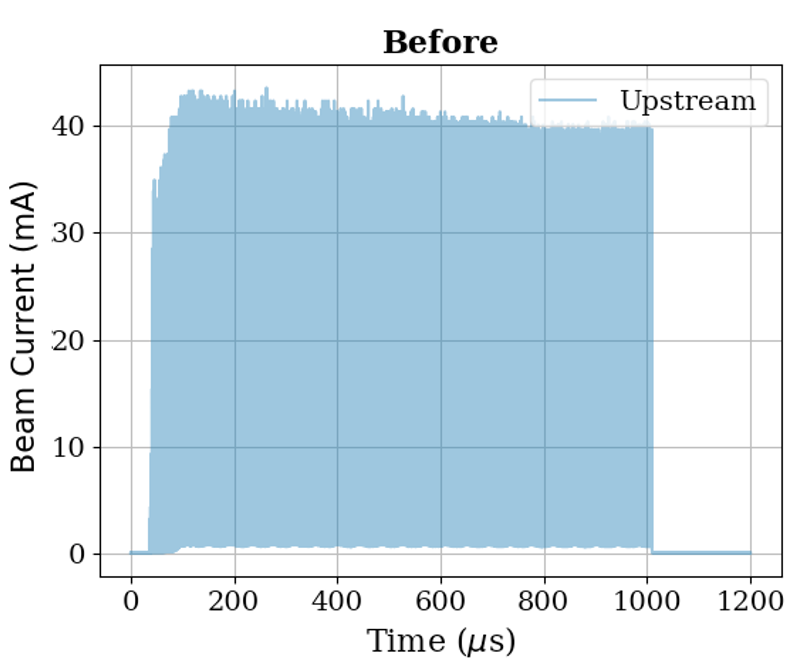}
\caption{Digitized trace of the beam current before the errant beam pulse}
\label{fig:sns-fulltrace}
\end{figure}

\subsection{Data Environment}
\begin{figure}[h]
    \centering
    \includegraphics[width=0.475\textwidth]{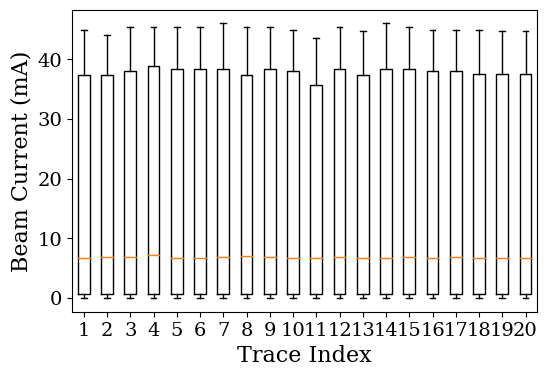}
    \caption{Box plot displaying random sample of 30 traces from the training set.}
    \label{fig:random_sample_normal}
\end{figure}

\begin{figure}[h]
    \centering
    \includegraphics[width=0.475\textwidth]{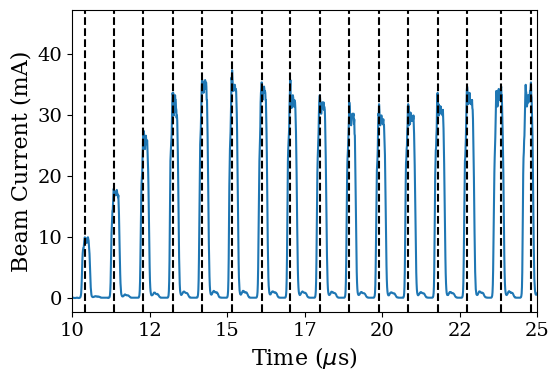}
    \caption{Sample of results using scipy.signal.find\_peaks function with height of 2 mA and distance of 75 samples between neighboring peaks.}
    \label{fig:peak_finding}
\end{figure}

A typical trace for training is shown in Figure \ref{fig:sns-fulltrace}. Each trace contains 120,000 samples, of which samples 3,000-13,000 are used for the Siamese model. 
Figure  \ref{fig:random_sample_normal} displays the box plots for 20 randomly selected traces from the normal set of traces. 
To remove statistically anomalous traces, each trace is required to have one peak every 750 ns.  
Peaks were identified using the find\_peaks method from the SciPy library~\cite{SciPy} with a minimum height of 2 mA and a distance of 75 samples (750 ns) between two neighboring peaks.  
A sample trace demonstrating the identified peaks is shown in Figure \ref{fig:peak_finding}.  
In addition, we required a minimum of 900  mini-pulses as setup for accelerator operations in order to exclude non-production setups.

To generate the data set for the Siamese model, we extracted normal and anomalous `Before' traces from the archived data from March of 2021 from the upstream sensor. From this we selected 4000 anomalous traces and compared each of them to 15 randomly selected normal traces. 
We used 4000 normal traces and also compared each to 15 randomly selected normal traces. 
The comparisons between normal traces are assigned a label of 0 and comparisons between normal and anomalous traces are assigned a label of 1. 
After applying the aforementioned data pre-processing, the data is divided into orthogonal training, testing, and validation data sets that contain 76800, 24000, and 19200 samples, respectively. 
The training, testing, and validation data sets contain equal numbers of normal to normal and normal to anomalous samples.
\subsection{Deterministic Siamese Model Results }
\begin{figure}[h]
\centering
\includegraphics[width=0.475\textwidth]{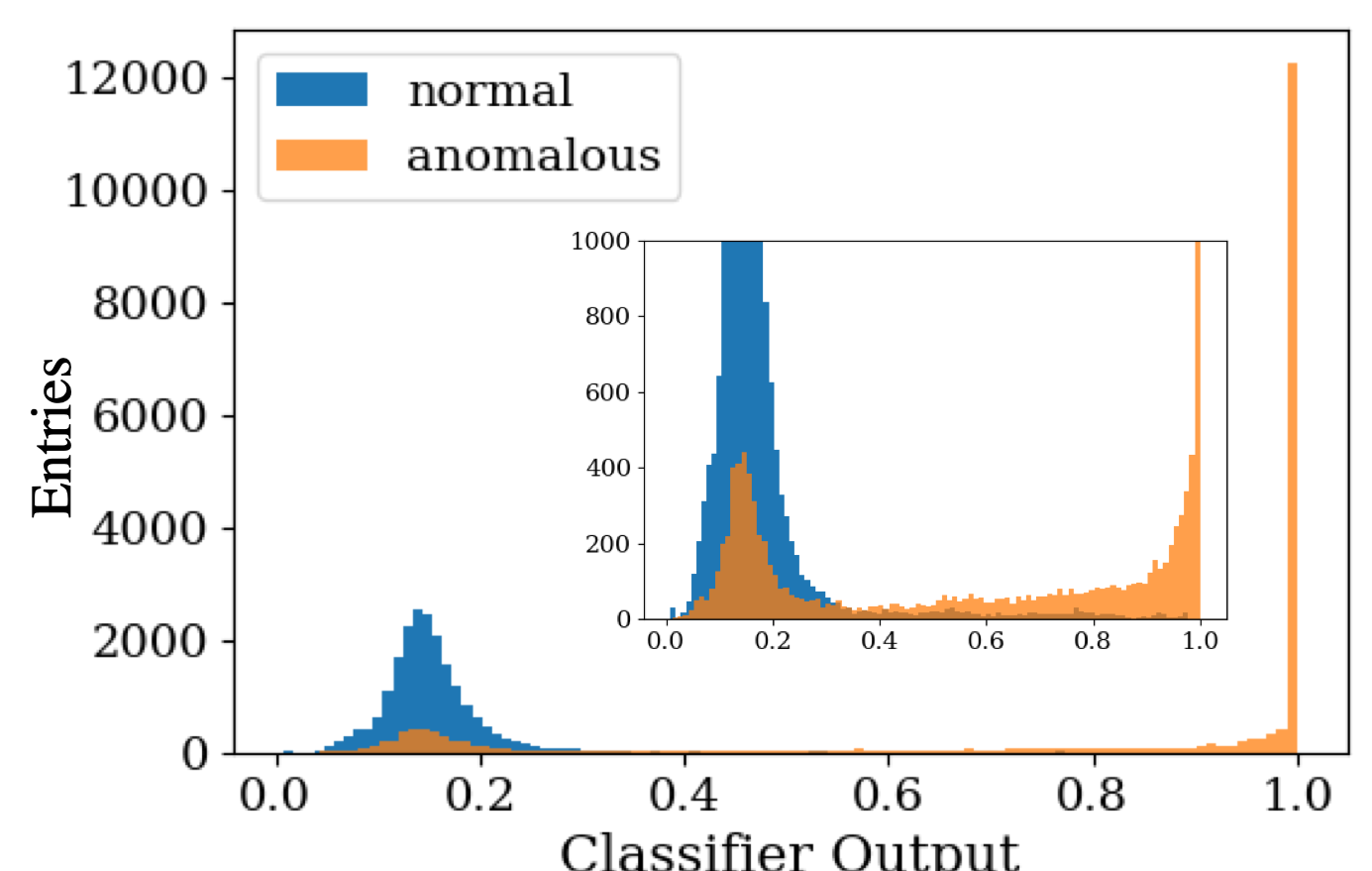}
\caption{Classifier output histogram (the sub-plot represents zoomed-in version of the same histogram plot)}
\label{fig:sns-output-histogram}
\end{figure}

Figure \ref{fig:sns-output-histogram} displays the results of the classification. The model identifies most of the anomalies with very high confidence while the remaining anomalies are misclassified as normal.
% \KISHAN{Classifier output separation figure}

\begin{figure}[ht]
\centering
\includegraphics[width=0.475\textwidth]{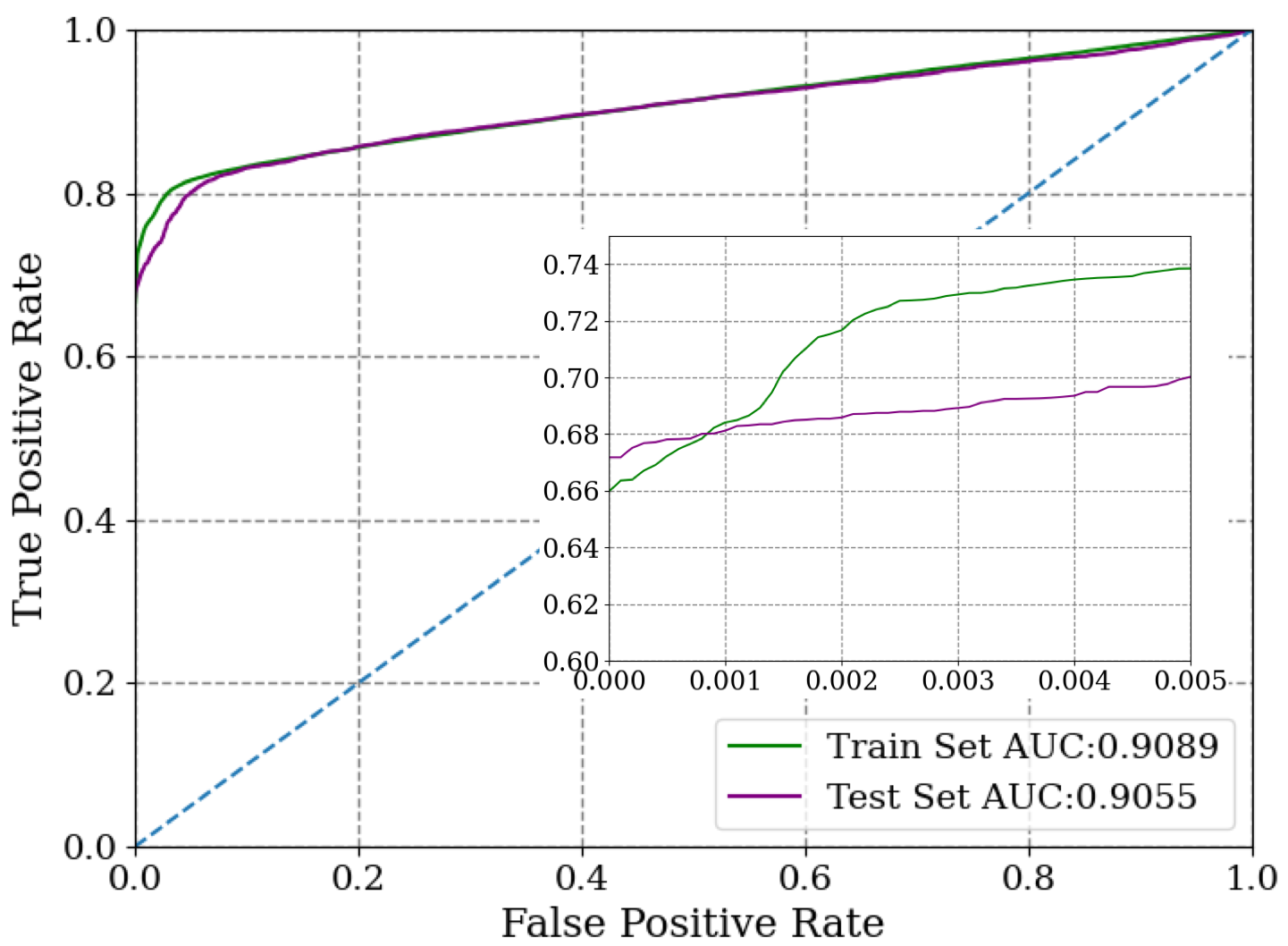}
\caption{Deterministic model ROC curves for train and test samples.}
\label{fig:sns-det-roc-curve}
\end{figure}

For the developed solution to be of practical use we must identify the maximum number of correctly identified anomalies while maintaining the FP rate below the established $0.05\%$.
As displayed in Figure \ref{fig:sns-det-roc-curve} (zoomed-in sub-plot), we have a true positive rate of more than $60\%$ on both train and test data sets.

% For the contrastive loss function we set the $\alpha$ term to 2 to assist the model to learn the similarities between normal pulses.
\subsection{Uncertainty Aware Siamese Model Results} 
To implement the uncertainties associated with the outputs of our Siamese model, we wrapped the last layer of the model with a Gaussian process layer. 
The uncertainty aware Siamese model not only provides a classifier output but also includes the uncertainty of the predictions.  We can explore how the model behaves in both dimensions.

\begin{figure}[ht]
\centering
\includegraphics[width=0.475\textwidth]{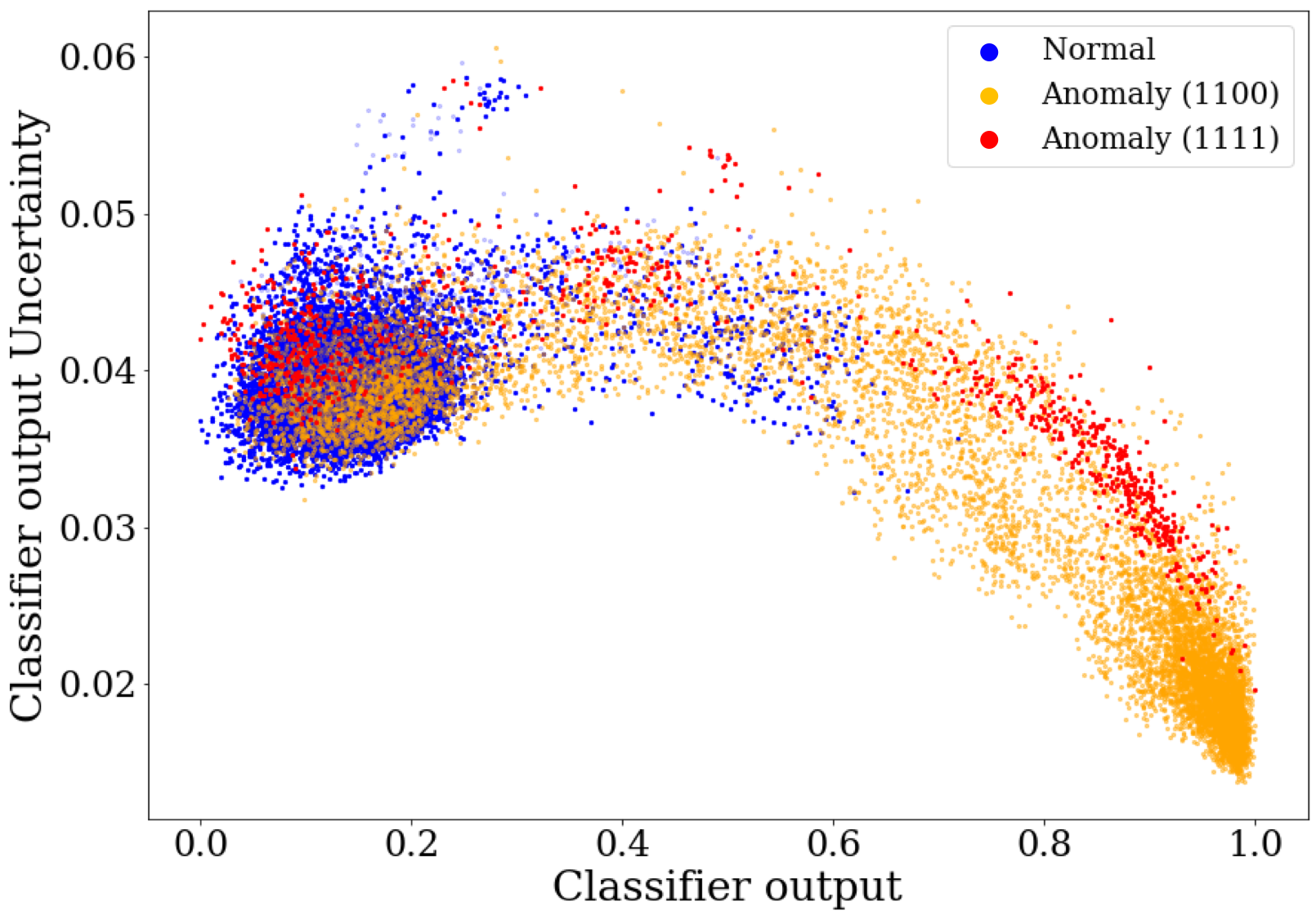}
\caption{Model predicted uncertainty vs uncertainty aware model prediction on a test data set. The blue dots are the normal pulses, the orange dots are the anomaly type used for training, and the red dots are the anomalous pulses that the model was not trained on.}
\label{fig:sns-error-scatter}
\end{figure}

As shown in Figure~\ref{fig:sns-error-scatter}, we introduced sample pulses (red dots) with anomalies that the model was not trained on.
We can see that the model can predict that these are anomalies with larger uncertainties meaning that the model is less confident in the predictions for previously unseen anomaly type.
The increase in uncertainty for these samples is consistent with our expectation.

\begin{figure}[h]
\centering
\includegraphics[width=0.475\textwidth]{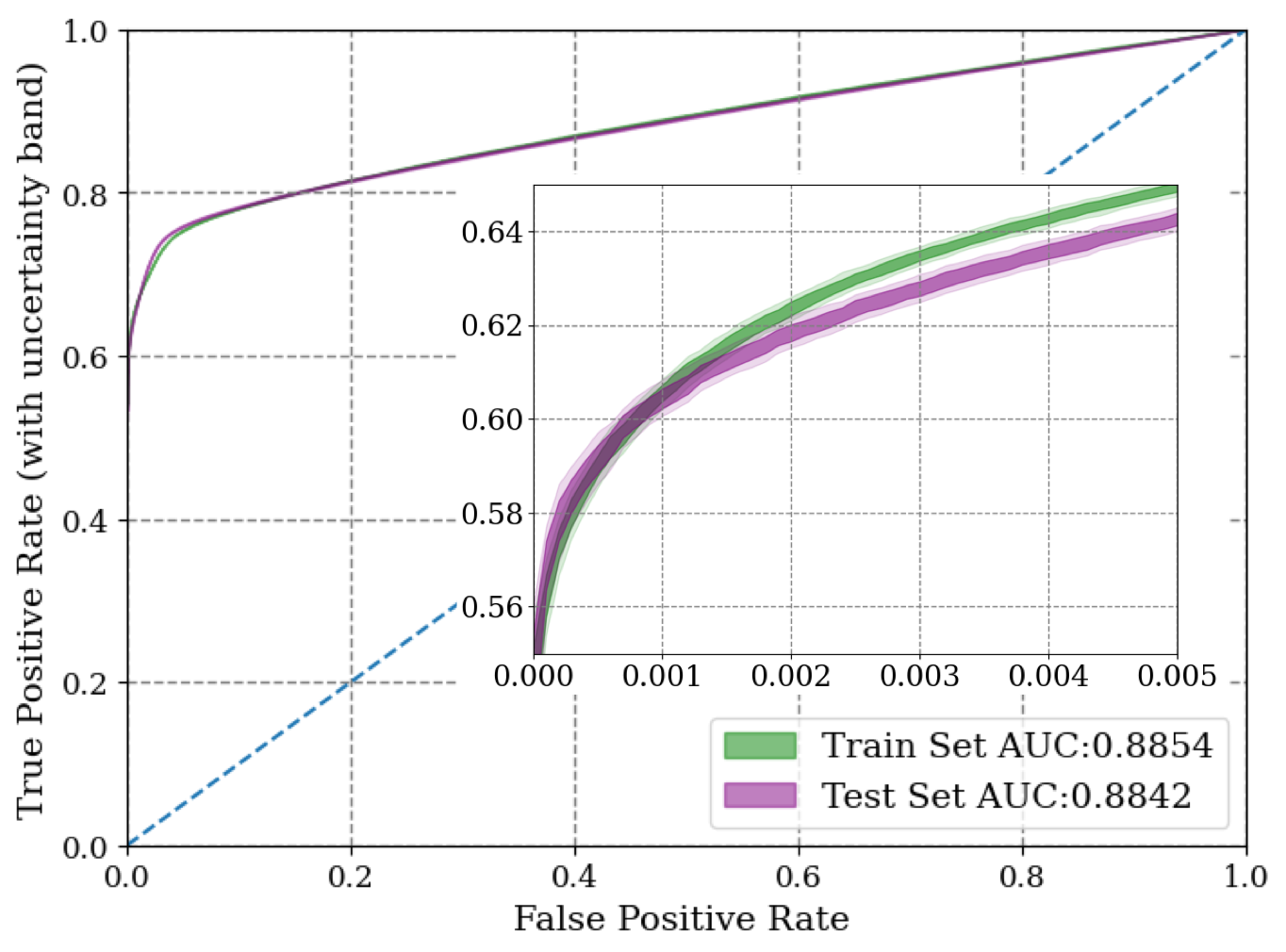}
\caption{Uncertainty aware model ROC curves for train and test samples. Dark error bands are for the $25\%/75\%$ quantiles and the light colour bands are for the $10\%/90\%$ values.}
\label{fig:sns-uq-roc-curve}
\end{figure}

In order to incorporate the model prediction uncertainties into the ROC curve, we smeared the model prediction output with its associated uncertainty using a Gaussian distribution. 
We conducted 250 trials to compute the ROC curve bands, as show in the Figure \ref{fig:sns-uq-roc-curve}. 
The dark bands represent range between $25^{th}$ and $75^{th}$ quantiles and the lighter bands represent the range between the $10^{th}$ and $90^{th}$ quantiles.

\begin{figure}[h]
\centering
\includegraphics[width=0.47\textwidth]{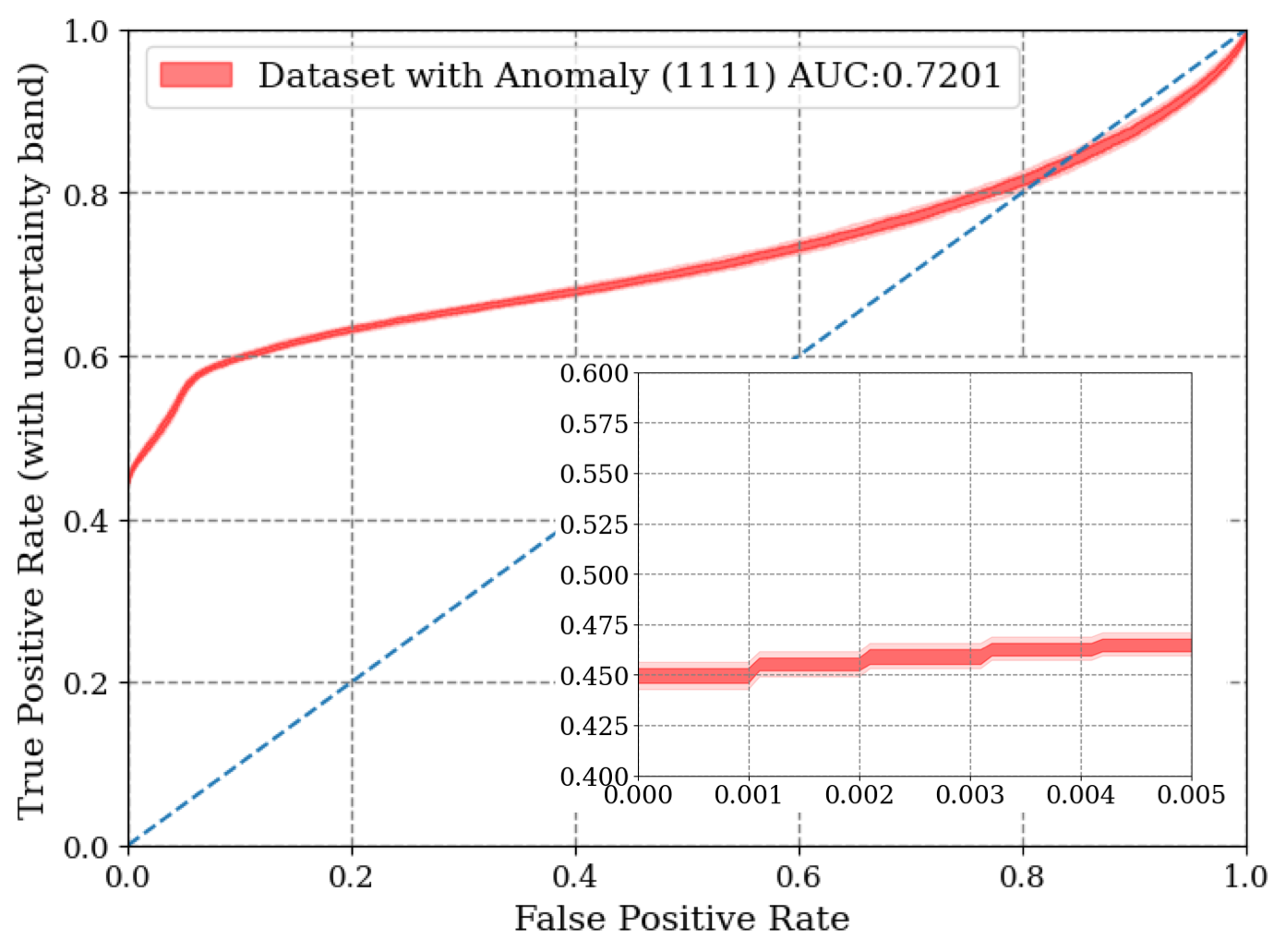}
\caption{ROC curve with uncertainty band for the inferences made on the anomaly type 1111 with the model trained on anomaly type 1100. The dark and lighter bands represents same range as Figure \ref{fig:sns-uq-roc-curve}.}
\label{fig:1111_roc}
\end{figure}

Figure \ref{fig:1111_roc} shows ROC curve uncertainty band for the unseen anomalies (type 1111) discussed above. Even though the model was not trained on these anomalies it is able to identify more than $45\%$ of the anomalies correctly while keeping the false positive below the threshold of $0.05\%$ though the predictions have higher uncertainties as can be seen in Figure \ref{fig:sns-error-scatter}. It shuld be noted that the scatter plot also shows that there is a threshold where the FP rate is neglectable.

\subsection{Siamese Inference CPU timing Results} 

The Siamese model has been tested on an Intel Core i9 with the timing results coming out at about an average of 2 ms per inference for the deterministic model and 4 ms for the distance aware model. 
The current DCM CPU is a Core i7 CPU and based on~\cite{CPUBench} it would still be able to infer within 10 ms. It should be noted that the DCM is scheduled to be upgraded to one of the Xeon CPU (W-2245 or E5-2618) listed in table \ref{tab:parameters}. 
As such, all CPUs considered are expected to easily complete the inference within the allotted time. 
As the data is transferred by DMA over a PXIe bus on a point by point basis as it is being sampled, the data is almost instantly available to the Real-Time CPU for processing. Current CPU usage is around 2-3 ms per 16.6 ms cycle, that means 16.6 - 1 - 3 or 12.6 ms of CPU time available to convert the data from fixed point to float, evaluate with the Siamese model, and send an abort signal back to the FPGA. We also tested the deterministic model inference running as a C++ code on NI PXIe-8840 (Core i5-4400E CPU) system with LabVIEW Real Time OS, one inference took under 4 ms.

%\begin{table*}[!htb]
%%\centering
%\caption{Expected model inference time for select computing hardware.}
%\begin{tabular}{ |p{2cm}||p{1cm}|p{1.95cm}|p{1.95cm}| } \hline
% \multicolumn{4}{|c|}{Performance} \\ \hline
% Intel CPU & CPU Mark Rating & Measured Deterministic/ Distance (ms) & Estimated Deterministic/ Distance (ms)\\ \hline
%Core i9-9880H&14,075&2.0/4.0&NA\\ 
%Xeon E5-2618Lv3&10,464&NA&2.7/5.4 \\
%Xeon W-2245&19,527&NA&1.4/2.9\\
%Core i7-5700EQ&5905&NA&4.8/9.5\\
%\hline
%\end{tabular}
%\label{table:timing}
%\end{table*}

\begin{table}[!htb]
%\centering
\caption{Expected model inference time for select computing hardware. Core i9-9880H and Core i5-4400E times were measured.}
\centering
\resizebox{\columnwidth}{!}{
\begin{ruledtabular}
\begin{tabular}{l|l|l}
 Intel CPU & CPU Mark  & Inference Time (ms) \\
&  Rating         &  Deterministic/Uncertainty         \\ \hline
Core i5-4400E & 3,251 & 3.2/NA (C++)\\
Core i7-5700  & 5,905   & 4.8/9.5\\
Core i9-9880H   & 14,075 & 2.0/4.0\\ 
%Xeon E5-2618Lv3 & 10,464 & 2.7/5.4 \\
Xeon E5-2618 & 10,464 & 2.7/5.4 \\
Xeon W-2245     & 19,527 & 1.4/2.9\\
%Core i7-5700EQ  & 5905   & 4.8/9.5\\

\end{tabular}
\end{ruledtabular}}
\label{tab:parameters}
\end{table}

%% file: Sections/Discussion.tex
\section{Discussion}

\subsection{Siamese Improvements}
% Distance preserving
In addition to the uncertainty quantification, we aim to implement a Class Activation Map (CAM) to highlight the distinct region(s) of the pulse the model focuses on when making a similarity classification. This method has been used extensively in recent years upon the realization that Convolutional Neural Networks can perform object localization without explicit supervision of the object \cite{CAM}. This can used to determine specific equipment failure classes. In the future, we can then compare these classes with failed equipment as indicated by the Machine Protection System (MPS) information.
\subsection{Implementation}
Given the performance of the method in both execution times and TP and FP rates, we plan to implement the model on the actual DCM system on its real-time system. In this paper we only analyzed the pulse immediately before the fault, however, the DCM also archives up to 25 preceding pulses. As such, we plan to study if the preceding pulses can also provide additional discriminating power which would allow us to identify fault earlier and more accurately. We plan to determine if different equipment failures have different durations from the first detection of an anomaly to the actual errant beam pulse and then apply the appropriate hold-off time for each different type of equipment failure. 

We now also have data becoming available from the Beam Position Monitors (BPM). The phase data from the BPMs is especially interesting as the phase relates directly to the momentum of the beam particles and the momentum is directly related to the acceleration process. Thus if the acceleration process is failing, we should see this in the phase data immediately. We hope that in this data even more precursors can be found.

\section{Conclusion}

In  this  paper,  we  have  described an uncertainty aware method to predict impending faults using data from a single data source. The FP rate can be set low enough that the performance of the accelerator is insignificantly affected while maintaining a TP rate high enough to benefit the accelerator. The method allows us to adjust our FP rate so that less beam pulse are wrongly aborted while still preventing over 40 percent of the beam loss events. Another practical aspect of the Siamese model is that we can also feed it pulses that we know are normal, no errant beam pulse either before or afterwards, and determine if this new normal pulse is still similar to the trained normal pulse. This will help us determine if we need to retrain the model. Execution times of the model are such that a practical implementation is possible which will halp us determine the benefits of the preventing errant beam.

\section{Acknowledgements}
The authors acknowledge the help from David Brown in evaluating Operations requirements , Frank Liu, for his assistance on the Machine Learning techniques, and Sarah Cousineau for making this grant work possible.

This manuscript has been authored by UT-Battelle, LLC, under contract DE-AC05-00OR22725 with the US Department of Energy (DOE). The Jefferson Science Associates (JSA) operates the Thomas Jefferson National Accelerator Facility for the U.S. Department of Energy under Contract No. DE-AC05-06OR23177. This research used resources at the Spallation Neutron Source, a DOE Office of Science User Facility operated by the Oak Ridge National Laboratory. The US government retains and the publisher, by accepting the article for publication, acknowledges that the US government retains a nonexclusive, paid-up, irrevocable, worldwide license to publish or reproduce the published form of this manuscript, or allow others to do so, for US government purposes. DOE will provide public access to these results of federally sponsored research in accordance with the DOE Public Access Plan (http://energy.gov/downloads/doe-public-access-plan).